\documentstyle[preprint,epsf,aps]{revtex}

\begin{document}
\title{A simple way to detect the state transition caused by the nondiagonal abelian Berry phase }
\author{Wang Xiangbin\thanks{email: wang@qci.jst.go.jp}, Matsumoto 
Keiji\thanks{email: keiji@qci.jst.go.jp}, Fan Heng\thanks{email: fan@qci.jst.go.jp}
 and Tomita Akihisa\thanks{email: a-tomita@az.jp.nec.com}
\\
        Imai Quantum Computation and Information project, ERATO, Japan Sci. and Tech. Corp.\\
Daini Hongo White Bldg. 201, 5-28-3, Hongo, Bunkyo, Tokyo 113-0033, Japan\\
Jian-Wei Pan\thanks{email: pan@ap.univie.ac.at}\\Institute fur Experimentalphysik, Universitat Wein, Boltzmanngasse 5, 1090 Wien, Austria }

\maketitle
\begin{abstract} 
In a nondegenerate system, the abelian Berry's phase will never cause transitions among the Hamiltonian's eigenstate.
However, in a degenerate system, it is well known that the state transition can be caused by the non-abelian
Berry phase. Actually, in a such a system, the phase factor is not always nonabelian. Even in the case that the phase is a
nondiagonal matrix, it still can be an abelian phase and can cause the state transitions. 
We then propose a simple scheme to detect such an effect of the Berry phase to the degenerate
states by using the optical entangled pair state. 
In this scheme, we need not control any external parameters during the quantum state evolution.
\end{abstract}
\newpage
Geometric 
phase\cite{panch,berry,simon} plays
an important role in quantum interferometry and many other disciplines. Many observable
quantum phyonemena are related to the geometric phase shift. The geometric
phase to  the degenerate states is qualitatively different from the one to the nondegenerate states, 
because it can cause the detectable effects which never happen to the nondegenerate states.
For example, to the degenerate states one can obtain the non-abelian geometric phase\cite{zee,mead,pines01,jan,jan2}. Such a non-abelian
Berry phase can cause transitions among the orthogonal states belonged the same eigenvalue of the Hamiltonian therefore
could be useful in quantum computation\cite{pachos}. 
Due to the non-abelian property, if we alternate two cyclic adiabatic operations, the states 
evolve differently. In this paper we will show that, to the degenerate states, there are {\it abelian }
phase factors
which are nondiagonal matrices.  This is to say, even the Abelian Berry phase
is qualitatively different from the one to the nondegenerate states, 
because it can cause  the transitions among the eigenstates 
after a  cyclic adiabatic evolution.  We also give a feasible
scheme to detect this type of nontrivil effect caused by the Berry phase to the degenerate states. 

The experimental testing
of geometric phase is an interesting and important topic. For a two level system, 
geometric phase is equal to half of the solid angle subtended
by the area in the Bloch sphere enclosed by the closed evolution loop of the eigenstate.
Most of the times, after certain time
evolution path, there are both dynamic and geometric phase shift, this causes extra treatment to
prove the effects of the geometric phase. 

Recently, applications of the geometric phase is extended the geometric quantum computation
\cite{pachos,ekert,vedral,xiaoguang,zana,duan,wang}, i.e., making the quantum gate through the geometric phase
shift. 
In particular, several of them are based on the idea of 
using the Berry phase\cite{pachos,zana,duan} shift to the degenerate states. Different from the case to
the nondegenerate states, to a degenerate system, the Berry phase after a cyclic adiabatic evolution may cause the bit flip or
the conditional bit flip between two states. This is rather significant because of its obvious
fault tolerance property.
For the nonabelian holonomic quantum computation, the computational space $C$ is always an
eigen space(spanned by a number of degenerate eigen states). The first clear
experimental proposal to produce the controllable Berry phase ship for the degenerate system
was raised by Duan et al\cite{duan}. 
The proposal suggests using trapped ions for the test. However, in this scheme, we will have to exactly
control the external parameters so that the eigenvalues of the instantaneous Hamiltonian keeps to be zero. 
In this paper, we propose  a simple scheme to test the states transition caused by 
Berry phase to the degenerate. Our scheme could be
more robust because  it requires little external parameters control in the whole process.  

Lets first see what is the non-trivil effect caused by the arbitrary Berry phase to the degenerate 
states.
Suppose orthonormal states $\{|\psi_i'\rangle\}$ spans a degenrate space of Hamiltonian $H$, i.e., 
all of the states are the eigenstates of $H$ with the same eigenvalue. 
If we change the Hamiltonian adiabatically and cyclically, there could be a non-trivil time evolution of the states in
the degenerate subspace. There is no dynamical contribution to the time evolution because the dynamical phase
is all zero. This time evlution operator  in general can be a matrix with nonzero off diagonal
elements.  Since there is no dynamical contribution, we name this time evolution operator as 
the geometric phase factor. Suppose after certain cyclic adiabatic process the Berry phase factor is $U$, 
i.e.
\begin{eqnarray}
\{|\psi_i'(t)|\rangle\}=U\{\psi_i'(0)\rangle\}.
\end{eqnarray}
Since $U$ is unitary, in principle we can always diagonalize it by another unitary matrix $S$ 
in the same degenerate space so that
$S^{\dagger}US$ is diagonalized. This is to say, in the basis $\{|\psi_i\rangle\}=S\{|\psi_i'\rangle\}$, 
the time evolution
operator is diagonal. Any state $|\psi_i\rangle=S|\psi_i'\rangle$ must be also an eigenstate
of $H$ due to the degeneracy property. And, all states of $\{\psi_i\}$ have the same eigenvalues 
with that of $\{\psi_i'\}$. 
Consequently, if any adiabatical process causes 
a diagonal time evolution(a phase shift)
in certain basis $\{|\psi_i\rangle\}$, the phase factor must be a non-diagonal matrix
in another basis $\{|\psi_i'\}$. This type of time evolution is in general regarded as {\it nonabelian}
and  may cause transitions among the
eigenstates, even though all  states keep on being the eigenstates of of the instantaneous 
Hamiltonian during the whole process. This transition never happens by the same operation to the 
nondegenerate eigenstates.  However, in certain case, the evolution operator can be {\it abelian} although
it is not diagonal. And as we will show in the following, such a nontrivil {\it abelian} phase can also cause state transitions.    
 
Consider the following entangled state for the spin half 
system
\begin{eqnarray}
|\Phi_{\pm}\rangle=\frac{1}{\sqrt 2}(|0_A1_B\rangle\pm|1_A0_B\rangle)\label{epr},
\end{eqnarray}         
subscripts $A$ and $B$ represent for the qubit $A$ or qubit $B$ respectively, 
0 is spin down and 1 is spin up.
Denoting $|\psi_1\rangle=|0_A1_B\rangle$ and $|\psi_2=|1_A0_B\rangle$. Obviously state $|\psi_1\rangle$ and $|\psi_2\rangle$ are dengenrate
in a $Z$-direction external field. 

Now we consider how to produce an {\it abelian} and {\it nondiagonal} matrix  phase factor to the two level degenerate states.
We can  first spatially separate spin $A$ from spin $B$. 
Obviously, we can also acquire a non-trivil Berry phase in the basis $|\Phi_{\pm}\rangle$ which is
linearly superposed by
state $|\psi_1\rangle$ and $|\psi_2\rangle$. 

We will show that we can have  a non-trivil effect if it is observed in
the $|\Phi_{\pm}\rangle$ basis through the abelian phase in the basis $|\psi_{1,2}\rangle$. 
Consider another set of states $|\Phi_{\pm}\rangle$, each of which is a linear superposed state of $|\psi_1\rangle$ and 
$|\psi_2\rangle$. We can first spatially separate qubit $A$ and $B$. Then rotate the local field
of qubit $B$ so that qubit $B$ evolve a $closed$ loop.
 After the evolution, the Berry phase for $0_A$ and $1_A$ are opposite and the value is propotional to
solid angle substended by the evolution loop at the degenerate point of the Bloch sphere for spin $B$. 
Explicitly, we have
\begin{eqnarray}
\Gamma_{0,1}=\pm \Omega/2
\end{eqnarray}  
where $\Gamma_{0,1}$ is the Berry phase shift for state $|0_B\rangle$ or $|1_B\rangle$ and $\Omega$ is the value of solid
angle substended by the evoultion loop. For example, if we rotate the field adiabatically around the $Z'$-axis
for a loop. Then we have $|\Gamma_{0,1}|=\pi \cos\theta$, $\theta$ is the angle between $Z$ and $Z'$. 
Without loss of generality, we use $\Gamma$ to denote  the  
Berry phase shift for state $|0_B\rangle$. At time $T$ when the evolution loop is completed
we have
\begin{eqnarray}
|\Phi(T)\rangle=\frac{1}{\sqrt 2}(e^{-i\Gamma}|\psi_1\rangle\pm e^{i\Gamma}|\psi_2\rangle\label{geo}).
\end{eqnarray}
Note that here no dynamical phase is involved because state $|\psi_1\rangle$ and state $|\psi_2\rangle$ are degenerate in the whole process.
Although the phase is diagonal in the $|\psi_{1,2}\rangle$ basis, it is $not$ diagonal in 
the $|\Phi_{\pm}\rangle$ basis. In such a non-trivil basis, the phase factor is given as
\begin{eqnarray}
\Sigma=\left(\begin{array}{cc} \cos\Gamma & -i\sin\Gamma\\ i\sin\Gamma & \cos\Gamma\end{array}\right)\label{geo1}
.\end{eqnarray}
This is to say, the time evolution is give as:
\begin{eqnarray}
\left(\begin{array}{c}|\Phi_+(T)\rangle\\ |\Phi_-(T)\rangle\end{array}\right)
=\Sigma \left(\begin{array}{c}|\Phi_+(T)\rangle\\ |\Phi_-(T)\rangle\end{array}\right).
\end{eqnarray}
Note that there is no dynamical contribution to this no-commutative phase factor $\Sigma$ because in the whole 
adiabatic process we have
\begin{eqnarray}
H(t)|\Phi_{\pm}(t)\rangle=0.
\end{eqnarray}
The accumation of the dynamical phase is trivilly zero.

This matrix phase factor $\Sigma$ is an {\it abelian} phase factor
although it can be non-diagonal. This is because all possible $\Sigma$ by the above operation  commutate with each other,
i.e., the matrices $\Sigma$ with different $\Gamma$ values commute. Although the phase factor here is abelian, it is different
from the usual Berry phase in the nondegenerate which is diagonal and which never causes transitions among the eigenstates
of the Hamiltonian.

Now we give a scheme to realize the above phase shift. It has been demonstrated extensively for the EPR state of eq(\ref{epr})
by the polarized photon pairs\cite{pan}. Now we are using polarized photons, state $|0\rangle$ and state $|1\rangle$ in eq(\ref{epr}) represent vertical or horizontal
polarizotion respectively. A polarized photon evolves by  the following time evolution operator when passing
through  a birefringence crystal  \cite{gisin}
\begin{eqnarray}
U(s)=e^{-i\hat\beta \cdot \sigma\cdot s}.
\end{eqnarray}
Here $s$ is the  the distance the light travels in the crystal, $\beta$ is the axis of the birefringence and $\sigma$
is the Pauli operator. Actually this is the spatial evolution operator.
We can produce a  non-abelian phase factor through the Berry phase shift
for state $|\Phi_{\pm}\rangle$ in the following way:

1. Prepare the polarized entangled pair stae $|\Phi\rangle=|\Phi_+\rangle$ or $|\Phi\rangle=|\Phi_-\rangle$ and let the two photons are flying in the opposite way.

2. Let photon $A$ fly in the medium $A$ made of a crystal whose birefringence axis is along $Z$-axis. 
Meanwhile, photon
$B$ is flying in the medium $B$ made up of the same crystal materials, 
however its birefringence axis changes gradually 
around $Z'$ axis for a loop. To meet the adiabatic condition we require the value ${\rm d} \hat \beta/ds$ very small. 
Mediums  $A$ and medium $B$ have the same length.
We set the length of the medium to be the value so that the refringence axis changes back to $Z$ direction. 
  
3. After the two photons comes out the crystal, 
the state $|\Phi_{\pm}\rangle$ is changed to the one by eq(\ref{geo1}). 

Note that here the evolution path of photon $B$ on the Bloch sphere can be any irregular loop. The Berry phase
is independent of the shape of the loop, it is only dependent on the area enclosed by the loop. 
In particular, if the birefringence $\hat \beta$ for medium $B$ is changed uniformly around $Z'$ axis and 
the angle between $Z$ and $Z'$ is set to  $\theta=\pi/3$, the output state is
$\Phi=\frac{1}{\sqrt 2}(|\psi_1\rangle-|\psi_2\rangle)$, which is othorgonal to the initial state. 
To improve the efficiency, we can also let photon $A$ also fly in a medium whose birefringence axis gradually
changes in a opposite manner as that of the medium $B$, i.e., polarization state of $A$ and $B$

In our schemes, we have used the EPR pair states. Obviously, we can also detect the non-abelian phase in the
basis defined by the non-perfect entangled state, namely $|\Phi_+\rangle=\cos\beta e^{-i\alpha}|01\rangle+
\sin \beta e^{i\alpha}|10\rangle$ and 
$|\Phi_-\rangle=\sin \beta e^{i\alpha}|01\rangle-\cos\beta e^{-i\alpha} |10\rangle$ where 
$\alpha$ and $\beta$ are non-zero real number 
satisfying the constraint $|\alpha|^2+|\beta|^2=1$.
 
In summary, we have proposed  simple schemes to produce and test 
the $abelian$ but {\it nondiagonal} Berry phase. It should be interesting to 
take a further study on realizing the  geometric controlled-NOT gate on the polarized photon qubits. 

{\bf Acknowledgement:} We thank Prof Imai for support. 
WXB thanks Prof N. Gisin, A. K. Pati and J. Pachos for fruitful discussions.

\end{document}